\documentclass[aip,preprint]{revtex4-1}

\usepackage{graphicx}
\usepackage{dcolumn}
\usepackage{bm}

\begin{document}


\title{Electric-field control of the  magnetic anisotropy \\ in an ultrathin (Ga,Mn)As/(Ga,Mn)(As,P) bilayer}

\author{T. Niazi}
\affiliation{Laboratoire de Photonique et de Nanostructures, CNRS, 91460 Marcoussis, France}
\author{M. Cormier}
\affiliation{Laboratoire de Photonique et de Nanostructures, CNRS, 91460 Marcoussis, France}
\author{D. Lucot}
\affiliation{Laboratoire de Photonique et de Nanostructures, CNRS, 91460 Marcoussis, France}
\author{L. Largeau}
\affiliation{Laboratoire de Photonique et de Nanostructures, CNRS, 91460 Marcoussis, France}
\author{V. Jeudy}
\affiliation{Laboratoire de Physique des Solides,
Universit$\acute{e}$ Paris-Sud -- CNRS, 91405 Orsay, France}
\affiliation{Universit$\acute{e}$ Cergy-Pontoise, 95000
Cergy-Pontoise, France.}
\author{J. Cibert}
\affiliation{Institut N$\acute{e}$el, CNRS -- Universit$\acute{e}$
Joseph Fourier,  38042 Grenoble, France}
\author{A. Lema\^{\i}tre}
\affiliation{Laboratoire de Photonique et de Nanostructures, CNRS, 91460 Marcoussis, France}%


\date{\today}

\begin{abstract}
We report on the electric control of the magnetic anisotropy in an
ultrathin ferromagnetic \mbox{(Ga,Mn)As}/\mbox{(Ga,Mn)(As,P)} bilayer with
competing in-plane and out-of-plane anisotropies.  The carrier
distribution and therefore the strength of the effective anisotropy is controlled by the gate voltage of a field effect device.
Anomalous Hall Effect measurements confirm that a depletion of
carriers in the upper \mbox{(Ga,Mn)As} layer results in the decrease of the
in-plane anisotropy. The uniaxial anisotropy field is found to
decrease by a factor $\sim 4$ over the explored gate-voltage range,
so that the transition to an out-of-plane easy-axis configuration is almost
reached.

\end{abstract}

\maketitle

The electric control of the magnetic state  is a very attractive
approach to bit-state manipulation in magnetic memories or logic
devices.  The first demonstrations in ferromagnetic semiconductors
were reported in ultrathin layers of \mbox{(In,Mn)As} or
\mbox{(Ga,Mn)As}. \cite{Ohno00, Chiba03, Stolichnov08}  In these
compounds, the ferromagnetic phase is carrier-mediated. The carrier
density is low enough to produce sizeable effects on the
ferromagnetic properties by depleting or accumulating carriers in a
field-effect device (FED). In the most recent studies, the electric
control of the magnetic anisotropy has been
demonstrated,\cite{Chiba08, Owen09, Sawicki10,Mikheev12} a first step toward
reversible magnetization switching by electrical means.\cite{Chiba10b}

However, achieving a significant modification of the magnetic
anisotropy by using its specific carrier-dependence \cite{Dietl01a}
requires a large change of the carrier density.\cite{Thevenard07a} Large depletion can be achieved in 4-7~nm thick \mbox{(Ga,Mn)As} layers.\cite{Chiba08,
Stolichnov08,Owen09} However, as the depleted region expands over the whole
ferromagnetic conductive channel, all its ferromagnetic properties are affected, which eventually leads to disappearance of the ferromagnetic order. Hence, it is an important
issue to find alternative methods to extend the magnetic
anisotropy tunability under electric field while preserving essential ferromagnetic parameters such as large Curie temperature and magnetization.

Here, we introduce and demonstrate a structure designed to enhance
the electrical control of the magnetic anisotropy without resorting
to excessively large depletion. This configuration eventually leads to the electric-field control of the in-plane and out-of-plane components of the magnetization.

\begin{figure}[]
\resizebox{0.90\columnwidth}{!} {\includegraphics{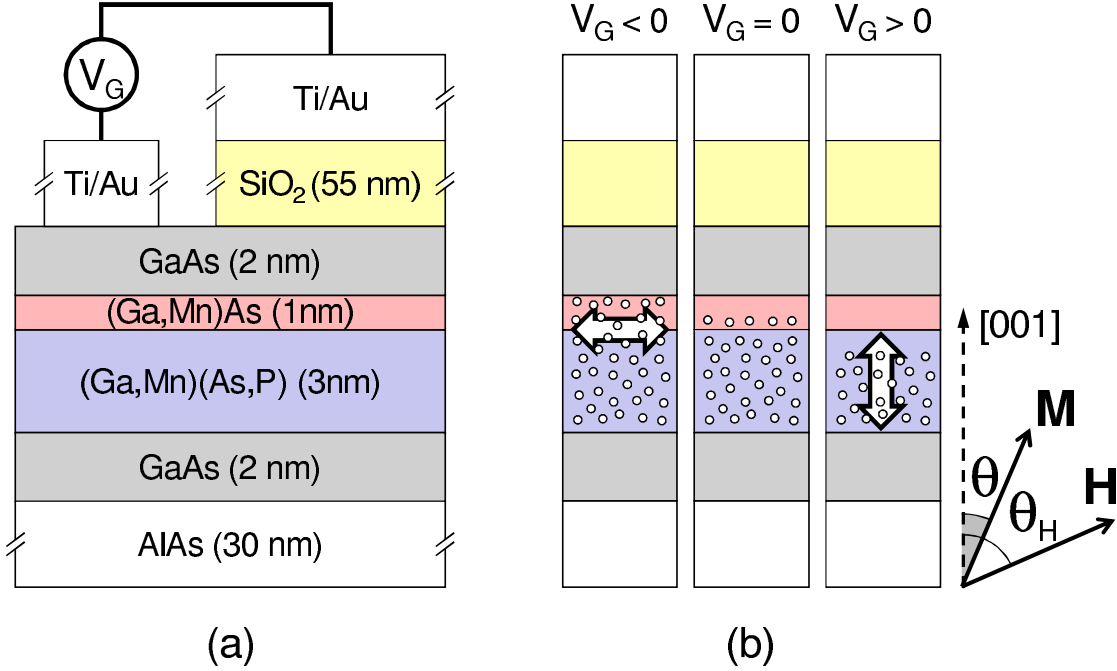}}
\caption{(a) Schematic diagram of the bilayer structure embedded in a
field-effect device. (b) A schematic drawing illustrating the principle of electric-field control of the bilayer easy axis (depicted by white arrows).}
\label{Schema}
\end{figure}

Our structure relies on an ultrathin
\mbox{(Ga,Mn)As}/\mbox{(Ga,Mn)(As,P)} bilayer embedded in a
metal-insulator-semiconductor FED (Fig.~\ref{Schema}a), grown on
(001) GaAs. In this specific strain configuration, the [001]
direction is an easy axis in \mbox{(Ga,Mn)(As,P)} and a hard axis
in \mbox{(Ga,Mn)As}.\cite{Lemaitre08} Therefore, the holes mediating
ferromagnetism explore two magnetic regions with competing
anisotropies. In our device, the bilayer thickness (4~nm) is smaller
than the exchange length ($\sim10$~nm),\cite{Haghgoo10} so that the
magnetization should be homogeneous over the hole gas region. Therefore,
the effective magnetic anisotropy results from the balance between
the contributions of the two layers. Applying a gate voltage
controls the depletion length, which affects mostly the layer closer
to the gate oxide, \mbox{(Ga,Mn)As} in our case. This should in turn lower
or increase the contribution of the \mbox{(Ga,Mn)As} layer to the
bilayer magnetic anisotropy (see Fig. \ref{Schema}b).

 The bilayer and the GaAs cap were grown at $\sim 220^\circ$C, the rest of the structure
at  $\sim 550^\circ$C. The as-grown Mn and P concentrations, both around 10~\%, were determined from calibration samples. The sample was annealed on a hot plate for 105~min at $200^\circ$C to decrease the amount of interstitial Mn ions, leading to an effective Mn moment concentration of about $\sim 5$~\% as measured on similar, but thicker samples.\cite{Cubukcu10}   Hall bars were processed by optical lithography and chemical etching. A 55~nm thick SiO$_2$ gate-oxide layer was
deposited by plasma-enhanced chemical vapor deposition at
$200^\circ$C. A Ti/Au film serving as the gate was then evaporated
over the conductive channel. The structural quality was checked, after annealing, by high-resolution scanning transmission electron microscopy. The magnetic layers does not exhibit any extended defects such as MnAs or MnP aggregates. High-angle annular dark-field (HAADF) Z-contrasted images revealed the bilayer structure: the \mbox{(Ga,Mn)(As,P)} layer appeared darker than the \mbox{(Ga,Mn)As} layer, with a  well defined interface. A thin layer ($\sim 2$~nm) consisting of gallium and arsenic oxide was observed, corresponding to the oxidization of the protecting GaAs cap layer.  Last, the SiO$_2$ layer appeared homogeneous, amorphous, showing no sign of crystallization, and stoichiometric as confirmed by energy-dispersive x-ray spectrometry.

Magneto-transport experiments were
performed on Hall bars aligned along $[110]$ (channel
length 215~$\mu$m, width 40~$\mu$m), connected by Ti (20~nm)/Au
(200~nm) metallic pads. The longitudinal and transverse
resistivities were measured by a lock-in technique with high-impedance pre-amplifiers, keeping the current low (20~nA) so that
the voltage drop across the Hall bar was negligible compared to the
gate voltage. The gate voltage excursion was limited by the breakdown voltage to $V_G=\pm 30$~V (corresponding to an electric
field of $\pm 5.5$~MV.cm$^{-1}$).

\begin{figure}[]
\resizebox{0.90\columnwidth}{!} {\includegraphics{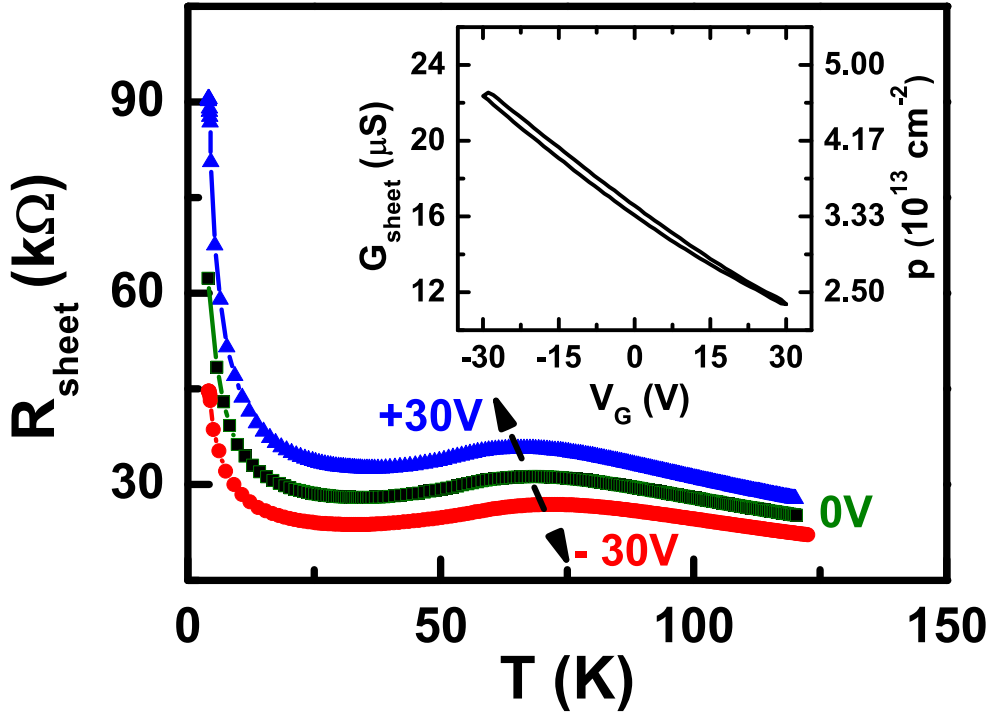}}
\caption{Temperature-dependence of the sheet resistance for three
values of the gate voltage, as indicated. Inset: sheet conductance
\textit{vs} $V_G$ measured at 4.2~K and zero magnetic field; the
right scale gives an estimate of the hole density.} \label{FigRV}
\end{figure}

The temperature-dependence of the sheet resistance
$R_\mathrm{sheet}$ is shown in Fig.~\ref{FigRV} for three values of
the gate voltage ($V_G=0, \pm 30$~V). Positive and negative voltages tend to achieve
the hole gas depletions schematically described in Fig.~\ref{Schema}(b).
Below 30~K, the resistance increases, indicating that the system is on
the insulator side close to the metal-insulator transition.  At
4.2~K, the sheet conductance $G_\mathrm{sheet}$ varies by a factor 2 for
$V_G=\pm 30$~V (Fig.~\ref{FigRV} inset). Hardly any hysteresis 
with electric field is observed, indicating the absence of carriers
deeply trapped within the oxide layer or at the oxide--semiconductor
interface. If we use the constant-mobility model of
Ref.~\onlinecite{Chiba08} and assume equal resistivities for both layers, we estimate a carrier density  $p
\sim 3.3\times10^{13}$~cm$^{-2}$ at 0~V and a density change $\Delta p\sim
2.3\times10^{13}$~cm$^{-2}$ between $\pm 30$~V (assuming a SiO$_2$ dielectric constant of 3.9). With these assumptions,
however, we probably overestimate $\Delta p$. Indeed, on the insulator side,
close to the metal-insulator transition, we may expect the presence
of weakly localized carriers, hardly contributing to the
conductance, but fully contributing to ferromagnetism and fully
affected by the electric field. Then, a better estimate of the
carrier density would be closer to the $\sim$10\% conductance variation measured at higher temperature.

The cusp around 60~K indicates the onset of the ferromagnetic phase.
A better estimate of the Curie temperature $T_C$ and its change
$\Delta T_C$ with $V_G$ can be inferred from the derivative of
$R_\mathrm{sheet}$ :\cite{Novak08} $T_C\sim 54$~K at $V_G=0$~V  and
$\Delta T_C \sim 3$~K between $\pm30$~V, a moderate change, comparable with that of previous reports.\cite{Chiba08, Stolichnov08,Owen09}

\begin{figure}[]
\resizebox{0.5\columnwidth}{!} {\includegraphics{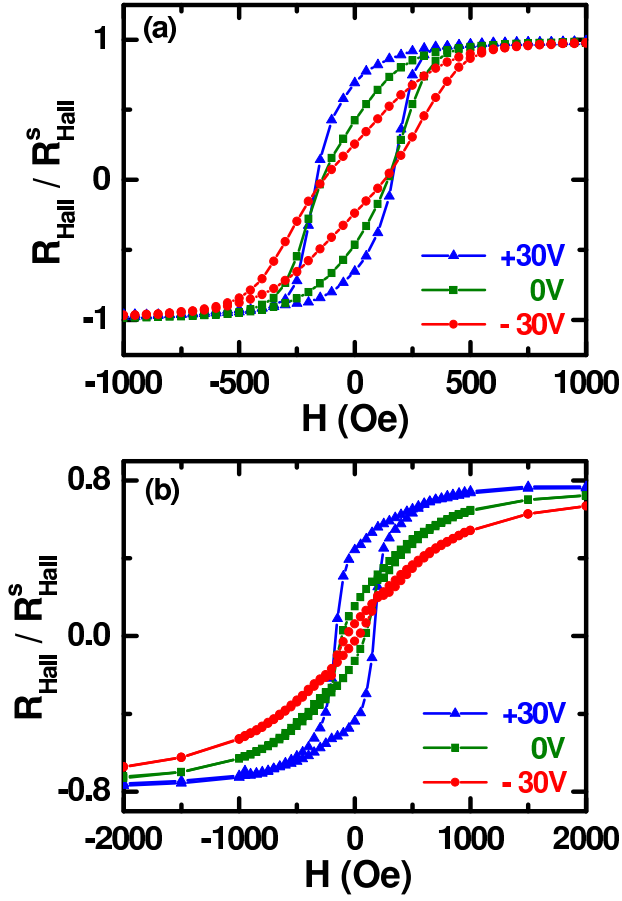}}
\caption{Field-dependence of the Hall resistance for $V_G=0, \pm
30$~V at 4.2~K. The field was applied along $[001]$ (a), or tilted
at $45 ^\circ$ towards $[1\bar{1}0]$ (b). Each curve is normalized
to the saturation value measured with the field along $[001]$.}
\label{FigRH}
\end{figure}

We now turn to the study of the magnetic anisotropy, deduced from 
anomalous Hall effect (AHE) measurements. Magnetic hysteresis loops measured for $V_G=0, \pm 30$~V are shown in Fig.~\ref{FigRH} for two directions of the
applied magnetic field, perpendicular to the sample
(Fig.~\ref{FigRH}(a)) and at $45 ^\circ$ towards $[1\bar{1}0]$
 (Fig.~\ref{FigRH}(b)). For each $V_G$-value, the Hall resistance
($R_\mathrm{Hall}$) was normalized to its value
($R_\mathrm{Hall}^s$) measured at saturation 
with the field along $[001]$ ($R_\mathrm{Hall}^s = 1.3$~k$\Omega$ at $V_G=0$~V). In (Ga,Mn)As and for the explored
magnetic-field range, $R_\mathrm{Hall}$ is dominated by the AHE:
$R_\mathrm{AHE} = \frac{R_S}{t} M_\perp$, where $R_S$ is the AHE
coefficient, $t$ the conducting layer thickness, and $M_\perp$ the
perpendicular component of the magnetization. $R_S$ usually depends
on the magnetoresistance. However, magnetoresistance was found to be
of the order of a few percent over the investigated magnetic-field range, sufficiently weak to consider that the field-dependence shown in Fig.~\ref{FigRH} mostly
reflects the variation of $M_\perp$. The small contribution of the
planar Hall effect was also not taken into account.

The curves in Fig.~\ref{FigRH}(a) display a saturation above
$\left|H\right| \sim 500$~Oe, and hysteresis below with a strong
shape modification  with $V_G$. In particular,
$\left|R_\mathrm{Hall}/R_\mathrm{Hall}^s\right|$ at $H=0$~Oe
decreases when $V_G$ is scanned from $+ 30$~V to $- 30$~V,
indicating a decrease of the remanent magnetization along $[001]$. As
nucleation of magnetization reversal and domain-wall propagation
contribute to the hysteresis-loop shape, we confirm the role of
anisotropy by considering loops obtained with the field applied at
$45 ^\circ$ towards $[1\bar{1}0]$, [Fig.~\ref{FigRH}(b)], which contain a reversible
contribution from coherent magnetization changes. During a field
scan from $2000$ to $500$~Oe,
$\left|R_\mathrm{Hall}/R_\mathrm{Hall}^s\right|$ decreases: this
indicates that [001] is a hard axis. The decrease is larger for
$V_G= - 30$~V than for $+ 30$~V, thus confirming that a negative
gate voltage reinforces the bilayer anisotropy in-plane character.

\begin{figure}[!t]
\resizebox{0.5\columnwidth}{!} {\includegraphics{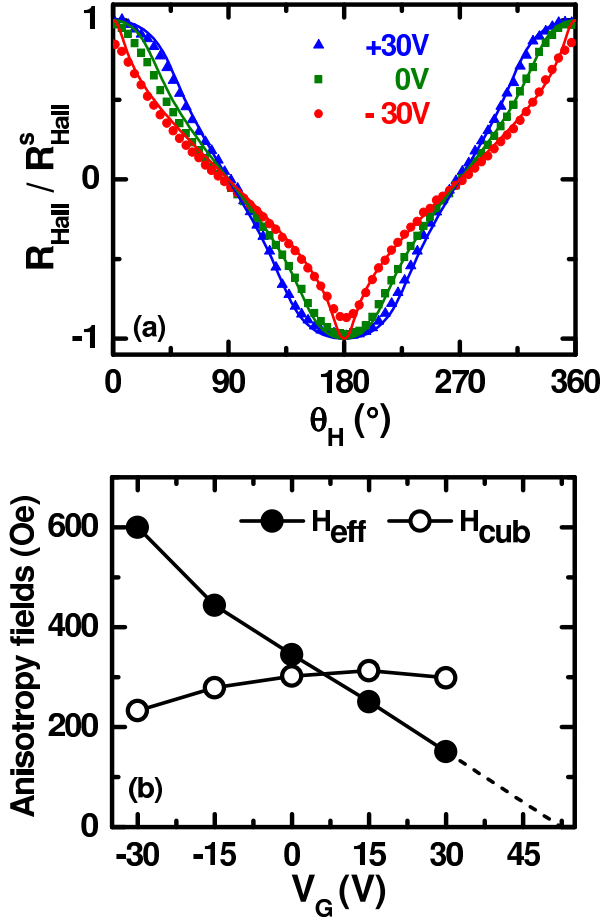}}
\caption{(a) Angular dependence of the Hall resistance for a 500~Oe field (symbols: experiment; lines:
fits with the model described in text). (b) Dependence of the anisotropy fields on the gate voltage, extracted from the fits at 500 and 1000~Oe.} \label{FigAngle}
\end{figure}

For a quantitative study, the applied field was rotated in the (110)
plane by an angle $\theta_H$ from [001]. For a field ($6000$~Oe)
much larger than the saturation field ($\sim 500$~Oe), the Hall
resistance exhibits a cosine-like variation (not shown): this is
compatible with a constant magnetization aligned on the
applied field, and $R_S$ almost independent of $\theta$.
The angular dependence for a field value close to the saturation
field is shown in Fig.~\ref{FigAngle}(a). Clear deviations from the
cosine law are observed, with sharper extrema around $\theta_H=0$ and
$180^\circ$ ($[001]$ and $[00\bar{1}]$ directions) at negative voltage.
We model this below as a result of the quasi-coherent magnetization rotation, with a competition between the applied magnetic field and the anisotropy.

The angular dependences measured at 500 and 1000~Oe, are fitted by
minimizing the free energy given by:
\begin{equation}
F(\theta)=M.\left( -H \cos(\theta-\theta_H)+\frac{H_\mathrm{eff}}{2} \cos^2\theta-\frac{H_\mathrm{cub}}{4}  \cos^4\theta\right)
\label{eq}
\end{equation}
where  $\theta$ is the magnetization angle. $H_\mathrm{cub}$
describes the cubic anisotropy. $H_\mathrm{eff}$ includes the
demagnetizing field and a combination of the magneto-crystalline,
uniaxial, and cubic anisotropy fields.\cite{Liu03} Eq.~\ref{eq} assumes that the magnetization remains in
the $(110)$ plane. This is likely since we find
$H_\mathrm{cub}<\left|H\right|$;  moreover an additional 
anisotropy within the (001) plane exists in (Ga,Mn)As which usually favors the $[1 \bar{1}0]$ orientation.\cite{Cubukcu10}  Figure~\ref{FigAngle}(a)  shows the good fit quality.

Figure~\ref{FigAngle}(b) shows the anisotropy fields extracted from the fits, as a function of the gate voltage. $H_\mathrm{cub}$ is found to vary weakly with $V_G$,
keeping a value ($\sim 300$~Oe) comparable to those reported in
Ref.~\onlinecite{Cubukcu10} for similar Mn concentrations. We
observe a strong variation of $H_\mathrm{eff}$. For each set of the anisotropy field values, the overall bilayer anisotropy can be understood from the free-energy landscape at $H=0$ (Eq.~\ref{eq}).  For $V_G<0$ and up to $\sim7$~V, $H_\mathrm{eff}>H_\mathrm{cub}$, so that there is a single free-energy minimum, corresponding to an easy axis  in the (001) plane. For $V_G \sim 7$~V, $H_\mathrm{eff}=H_\mathrm{cub}$, hence a second minimum appears corresponding to the [001] direction. At $V_G=30$~V, $H_\mathrm{eff}/H_\mathrm{cub}
\sim 1/2$, hence we reach the point where both minima are
equivalent. By extrapolating the data in Fig.~\ref{FigAngle}(b), a single free-energy minimum corresponding to an easy axis along the [001] direction would be obtained
at $V_G \sim 50$~V. Over such a voltage excursion, the magnetic easy axis could be tuned from in-plane to out-of-plane, achieving the magnetic configuration depicted in Fig.~\ref{Schema}(b).

Our results are consistent with an electric control of the
competition between the opposite uniaxial anisotropies of each
layer [Fig.~\ref{Schema}(b)]. At $0$~V,  the bilayer has a moderate in-plane anisotropy, showing
that the topmost (Ga,Mn)As layer is not totally depleted by transfer
to the hole traps at the interface with SiO$_2$.
\cite{Chiba08, Sawicki10} This suggests a rather moderate trap density. We cannot also exclude a carrier transfer from the \mbox{(Ga,Mn)(As,P)} layer toward the \mbox{(Ga,Mn)As} layer, as P substitution induces a valence-band offset between the two magnetic layers. At $-30$~V, the depletion length is reduced, the hole gas
now extends more into the \mbox{(Ga,Mn)As} layer, and the in-plane anisotropy
 is reinforced. At $+30$~V, depletion reduces the
contribution from the \mbox{(Ga,Mn)As} layer, enhancing the relative contribution of the \mbox{(Ga,Mn)(As,P)} layer to the bilayer anisotropy. The fact that we did not reach
an out-of-plane easy-axis at $+30$~V suggests that the depletion
induced by the gate is  rather weak, smaller that what is
suggested by the constant-mobility model applied down to low
temperature.

Other origins for this electrical tuning of the anisotropy can be ruled out. Although the gate voltage is expected to shift the boundary between the hole gas and the depleted region, as schematized in
Fig.~\ref{Schema}, this boundary is not totally abrupt, and a
variation of the 3D carrier density is expected, at least when
approaching total depletion of the channel. This mechanism has been used to
electrically control the Curie temperature or the cubic anisotropy.\cite{Chiba08, Owen09, Sawicki10,Mikheev12}
In the present case, these two parameters vary only very weakly with the gate voltage, so
that we can rule out such a mechanism for the variation of
$H_\mathrm{eff}$. A weakening of the contribution from the demagnetizing
effect is also not likely, since we worked far below the Curie temperature.

In conclusion, clear modifications of the magnetic anisotropy have
been achieved in an ultrathin \mbox{(Ga,Mn)As}/\mbox{(Ga,Mn)(As,P)} stack upon
applying  electric field. We attribute this effect to the electric
control of the competition between the in-plane and out-of-plane anisotropies of both layers. While in this study, large electric fields
were needed to observe sizeable modifications, switching to
high-\textit{k} gate oxides, such as HfO$_2$, should allow a better
control of the bilayer magnetic anisotropy with lower values of the
gate voltage.

The work was performed in the framework of the ANR MANGAS project (2010-BLANC-0424).

%
%
\end{document}